# Introducing Novel Planar Micromixers with Pillars and Gaps and Studying the Impact of Various Geometric Parameters on the Efficiency of Micromixers

Ali Kheirkhah Barzoki[1,*]

[1] Department of Mechanical Engineering, Sharif University of Technology, Tehran, Iran

[*] Correspondence: ali_kheirkhahbarzoki@mech.sharif.edu

## Abstract

Chemical bioreactions play a significant role in many of the microfluidic devices, and their applications in biomedical science have seen substantial growth. Given that effective mixing is vital for initiating biochemical reactions in many applications, micromixers have become increasingly prevalent for high-throughput assays. In this study, numerical study is conducted to examine the fluid flow and mass transfer characteristics in novel micromixers featuring an array of pillars. The study explores the effects of pillar array design on mixing performance and pressure drop, drawing from principles such as contraction-expansion and split-recombine. Two configurations of pillar arrays are introduced, each undergoing investigation regarding parameters such as pillar diameter, gap size between pillar groups, distance between pillars, and vertical shift in pillar groups. Subsequently, optimal micromixers are identified, exhibiting mixing efficiency exceeding 99.7% at moderate Reynolds number ($Re = 1$), a level typically challenging for micromixers to attain high mixing efficiency. Notably, the pressure drop remains low at 1102 Pa. Furthermore, the variations in mixing index over time and across different positions along the channel are examined. Both configurations demonstrate short mixing lengths and times. The combination of rapid mixing, low pressure drop, and short mixing length positions the novel micromixers as highly promising for microfluidic applications.

## Keywords

Micromixer, Mixing, Pillar, Obstacle, mass transfer, microfluidic, FEM,

## 1. Introduction

The concept behind microfluidic devices is to downsize a traditional laboratory into a miniaturized chip, incorporating various functions such as detection[1,2], mixing[3,4], synthesis[5–7], and separation[8–10] at the micro- and nano-scale. These devices are extensively used in chemical analysis[11,12], chemical synthesis[13,14], and biomedical analysis[15] due to their speed, accuracy, and minimal reagent requirement. Efficient mixing holds significant importance in these devices, making the attainment of quick and uniform mixing with minimal fabrication complexity a key design objective. Extensive research conducted over time has resulted in the introduction of diverse micromixer designs.

Microfluidic flow in micro-channels is typically laminar due to their small length scale, characterized by a low Reynolds number (Re). In the co-flow of two homogeneous and miscible fluids within a channel, mass transport is mainly governed by molecular diffusion. This makes microfluidic mixing particularly challenging. This situation exacerbates in applications where the mixing involves large analytes such as DNA or nanoparticles. One approach to microfluidic mixing involves the use of droplets to mix reagents in the process of droplet formation. When two immiscible fluids meet at a junction, various forces, such as viscosity, surface tension, and pressure gradient, come into play, causing the droplet to separate from the dispersed phase and flow downstream. By maintaining a controlled droplet size, this method prevents axial dispersion of mixing components, allowing for swift mixing through the internal circulation of the droplet[16,17]. Belousov et al. introduced an asymmetric flow focusing droplet generator to enhance mixing during the droplet formation stage, demonstrating a six-fold increase in mixing speed compared to a symmetric design[18]. Several studies have concentrated on improving the mixing index in microfluidic systems based on droplets by integrating micromixers immediately following the droplet generator[19–23]. The incorporation of micromixers induces advection within the droplets, consequently enhancing the mixing process. Nonetheless, droplet microfluidics exhibits heightened sensitivity to the conditions governing droplet formation, requires more intricate instrumentation, and introduces certain limitations related to the compatibility of materials with oils and surfactants.

Another promising approach is the utilization of micromixers to enhance the mixing of analytes. In the last two decades, there has been a surge in the design and study of micromixers[4,17,19,23–26]. These micromixers generally fall into two main categories: active mixers, which enhance mixing through external forces that disturb and stir the fluid[24,25,27–29], and passive mixers, where convective re-circulations and vortices induce folding and extension of the fluid–fluid interface[23,26,30–32]. This results in a reduction in the length-scale minimizing the distance over which molecular diffusion needs to act for complete mixing. Active approaches can achieve high mixing performance but are often more complex and expensive to integrate and fabricate compared to passive mixers. On the other hand, passive mixers can also achieve a high degree of mixing, but they may require significant pressure to drive flow and need a simple structure to facilitate fabrication and prevent channel clogging.

The performance of passive micromixers is inherently influenced by the channel geometry. A common strategy for achieving mixing in a microchannel involves employing a series of repetitive mixing units. Each mixing unit represents a geometric structure specifically designed to elongate the fluid interface through fluid–structure interaction. Several studies have concentrated on the design of various serpentine channels aimed at inducing secondary flows and augmenting mixing efficiency[20,26,31,33–36]. An alternative strategy for improving mixing efficiency involves incorporating obstacles into the channel[37,38]. The presence of obstacles in the channel leads to an extended fluid-fluid interface, resulting in increased mixing as the fluids traverse these impediments. In certain studies, researchers have integrated serpentine channel designs with obstacles, combining the advantages of both approaches[24,35,39–41]. Introducing obstacles into the channel may result in an elevated pressure drop, but it presents the advantage of enabling proper mixing of reagents within a relatively compact length. Conversely, serpentine designs can elongate the channel length and introduce complexity to the overall design. Consequently, devising an optimal and straightforward arrangement of obstacles holds significant promise.

In this study, we explored the mixing of two co-flowing fluids through the inertial flow deformation induced by a series of two-dimensional cylindrical obstacles (pillars). We examined two distinct configurations, manipulating parameters such as pillar diameter, gap size between pillar groups, distance between pillars, and the vertical shift between groups of pillars. Both configurations showed the potential for achieving a mixing index higher than 99.7% within a relatively short channel length and time making them applicable to diverse applications, particularly those requiring a compact design.

## 2. Mathematical Model

### 2.1. Governing Equations

The mixing analysis involves two key principles: mass transport and momentum transport. To obtain the flow field for incompressible fluids, the conservation of mass (continuity) and the conservation of momentum (Navier-Stokes equations) must be solved. Considering laminar flow ($Re \approx 1$), steady state, and incompressible fluid, these equations are outlined below:

$$\nabla . \vec{u} = 0 \tag{1}$$

$$\rho(\vec{u} . \nabla)\vec{u} = -\nabla P + \mu \nabla^2 \vec{u} \tag{2}$$

where $\vec{u}$ denotes the velocity vector ($m/s$), $\mu$ represents the dynamic viscosity ($Pa.s$), $P$ shows the pressure ($Pa$), and $\rho$ denotes the fluid density ($kg/m^3$).

To model the mixing phenomenon, the convective-diffusive mass transport equation was employed:

$$\partial c / \partial t + \vec{u} . \nabla c = D \nabla^2 c \tag{3}$$

Here, D represents the diffusion coefficient ($m^2/s$) and c denotes the dye concentration ($mol/m^3$). To determine the mixing index (MI), the dye concentration in each mesh cell was taken into account, employing the following formula[42]:

$$MI\ (\%) = \left(1 - \sqrt{\frac{\iint (c-\bar{c})^2 dA}{A . \bar{c}^2}}\right) \times 100 \tag{4}$$

where, c and $\bar{c}$ represent the dye concentration and average dye concentration, respectively. A denotes the area in which the mixing index is calculated. MI varies between 0 to 100%. $MI = 0$ corresponds to no mixing, while $MI = 100\%$ means complete mixing. As the MI increases, the mixing efficiency is improved, and the concentration of species becomes more uniform. In this study, the mixing index was calculated in a specific area at the end of the channel, called mixing box (see **Fig. 1A**).

## 2.2. Boundary Conditions

The study on fluid flow in the microchannel implemented the no-slip boundary condition on the channel walls, while maintaining a constant zero gauge pressure at the outlet. For the inlets, constant flow rate (25 $\mu l/min$) with a parabolic velocity distribution was considered. Considering the characteristic length of the inlet, Re was obtained 0.83. Deionized (DI) water, with a density of 1000 $kg/m^3$ and a dynamic viscosity of 0.001 $Pa.s$, was chosen as the working fluid. To be noted that since the dye solution was considered to be diluted, the density and viscosity of the solution was considered as DI water. For the mass transfer investigation, the walls were subjected to a no-mass flux boundary condition, and constant concentrations of 0 and 1 $mol/m^3$ were set at the inlets. A diffusivity constant of $5.75 \times 10^{-10}\ m^2/s$ was selected.

## 2.3. Numerical Model

The assessment of mixing efficiency, which measures the uniformity of various analytes, was conducted through numerical simulations employing the finite element method (FEM). This process involved the simultaneous solution of two sets of equations. The fluid flow equations were addressed using the Parallel Direct Sparse Solver (PARDISO) with a residual tolerance of 1E-3, while the mixing equation was solved using the Multifrontal Massively Parallel Sparse Direct Solver (MUMPS) with left preconditioning and a residual tolerance of 1E-3. First-order and second-order elements were utilized for discretizing pressure and velocity, respectively, and linear discretization was applied to the mixing study. The Newton method was employed for the linearization of non-linear equations at each time step. The determination of the time step ($\Delta \tau$) relied on the Courant number ($Co = 0.25$) to accommodate unsteady simulations (**Eq. 5**). In this equation, u and $\Delta h$ denote the average fluid velocity and mesh size, respectively. Triangular (tri) elements were employed for discretizing the entire domain, encompassing boundary layers. Although three-dimensional (3D) geometries are generally more precise, 2D geometries were chosen for the simulations due to their ability to provide valuable information and maintain acceptable consistency with experimental data without necessitating extensive computational resources[18,43–45].

$$Co = u\Delta\tau/\Delta h \tag{5}$$

## 2.4. Geometry of Micromixers

In this research, the microchannel features a straight design with an array of pillars. As depicted in **Fig. 1**, there are two primary pillar configurations: slanted configuration (**Fig. 1A**) and arrowhead configuration (**Fig. 1B**). The system includes two fluid inlets with dye concentrations of 1 and 0 $mol/m^3$. It is important to note that all properties of the two fluids are identical, except for the dye concentration. The channel has an outlet at its end, and a mixing box is utilized at the end of the channel to compute the mixing index within it (**Fig. 1A**).

In both configurations, there are some groups of pillars, each consisting of four rows (see **Fig. 1A**). These groups are separated by a specific gap size (G), which is calculated as $S_x$ - $S_a$ (**Fig. 1B**). Within each pillar group, the spacing between pillars in each column (**Fig. 1A**) is determined by two fixed parameters, $\Delta x$ and $\Delta y$ (**Fig. 1B**). The parameter $S_a$ establishes the distance between the columns of pillars in each group, with its values specified in **Table 1**. The distance between separate groups of pillars is determined by $S_x$ and $S_y$. As previously mentioned, $S_x$ equals $S_a + G$. If G equals 0, there will be no gap between the groups of pillars, resulting in a continuous array through all the channel. **Table 1** provides the geometrical parameters and their values. To be noted that $D_p$, $S_a$, $S_y$, and G get various values and the effect of changing each parameter on the mixing index is studied in this research.

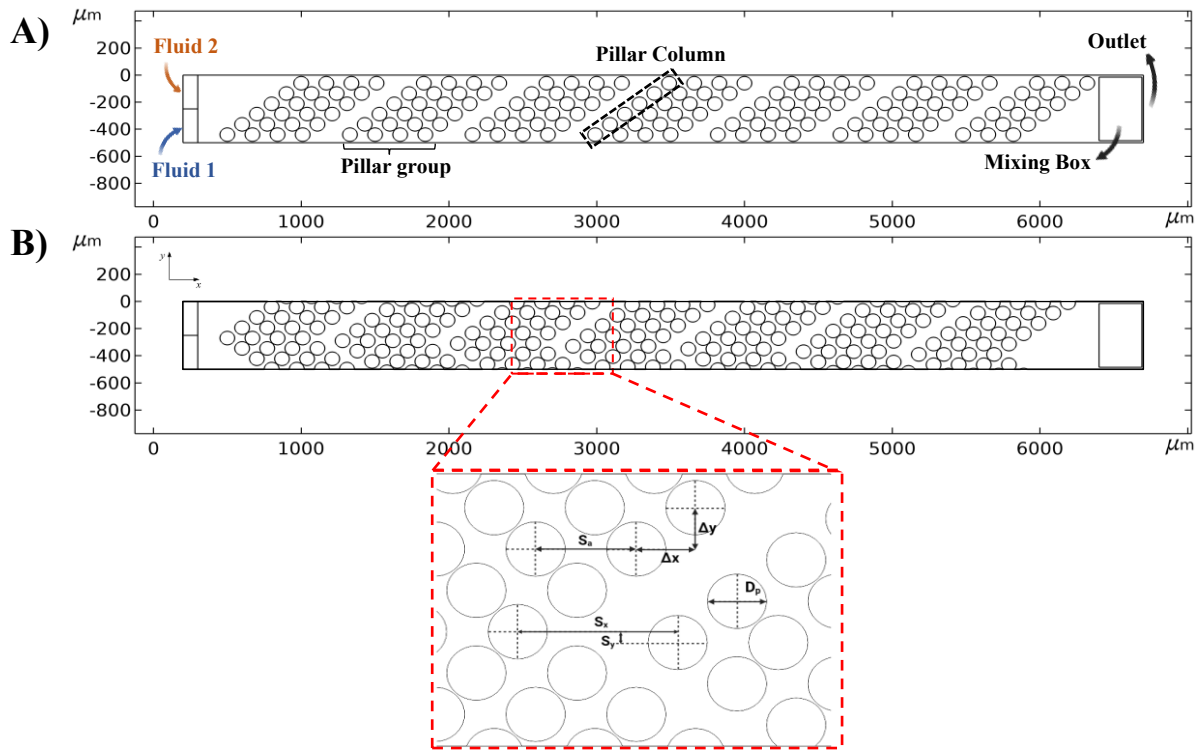

**Figure 1. Sketch of the Geometries**. **A)** Slanted configuration and flow path, and **B)** arrowhead configuration and geometrical parameters.

**Table 1. Geometric dimensions of the micromixers**.

| Parameters | $D_p$ | $\Delta x$ | $\Delta y$ | $S_a$ | $S_x$ | $S_y$ | $G = S_x - S_a$ |
|---|---|---|---|---|---|---|---|
| Values ($\mu m$) | 50 | 100 | 76 | 170 | $S_a+0$ | 10 | 0 |
| | 75 | | | 190 | $S_a+50$ | 20 | 50 |
| | 100 | | | 210 | $S_a+100$ | 30 | 100 |
| | 125 | | | 230 | $S_a+150$ | 40 | 150 |
| | | | | 250 | $S_a+200$ | | 200 |

## 2.5. Mesh Independence Study

As previously stated, the study examines various configurations of pillars in a rectangular channel to assess mixing efficiency. To ensure reliable and consistent numerical simulation results, each microchannel must have a designated mesh configuration containing an appropriate number of grids. Consequently, a thorough mesh independence study is conducted for the slanted configuration with $G = 0$, $S_a = 250\ \mu m$, and $D_p = 100\ \mu m$, and the corresponding results are detailed in **Table 2**. The outcomes reveal an improvement in the mixing index when the element size is reduced to values below 20 μm. Although 20, 15, and 7 μm show comparable results, the selection of 15 μm is favored for its advantages in terms of both reduced computational time and high accuracy. Consequently, the subsequent simulations adopted a mesh element size of 15 μm.

**Table 2. Mesh independence study**

| Element Size in $\mu m$ | Number of Elements | Mixing Index (%) |
|---|---|---|
| 50 | 12178 | 94.09 |
| 30 | 12293 | 89.93 |
| 25 | 14521 | 86.51 |
| 20 | 20314 | 64.02 |
| 15 | 32538 | 63.81 |
| 7 | 149944 | 62.79 |

## 2.6. Model Validation

To validate our computational model, we utilized the results from a previous experimental study conducted by Xia et al.[46] featuring obstacles and gaps, as in this study. For this purpose, one of their geometries with suitable boundary conditions and mesh configuration, as shown in **Fig. 2A, B**, was simulated. **Fig. 2C** illustrates the numerical mixing index results obtained through our computational model, along with the experimental and numerical results by Xia et al.[46]. It is evident that the numerical results surpass the experimental data, attributed to certain assumptions made in the simulation. Across the range of Re = 0.1, 1, 10, 20, 40, and 60, the mixing index progressively increases with Re, demonstrating a consistent correspondence between the mixing index results from our computational model and those reported by Xia et al.[46].

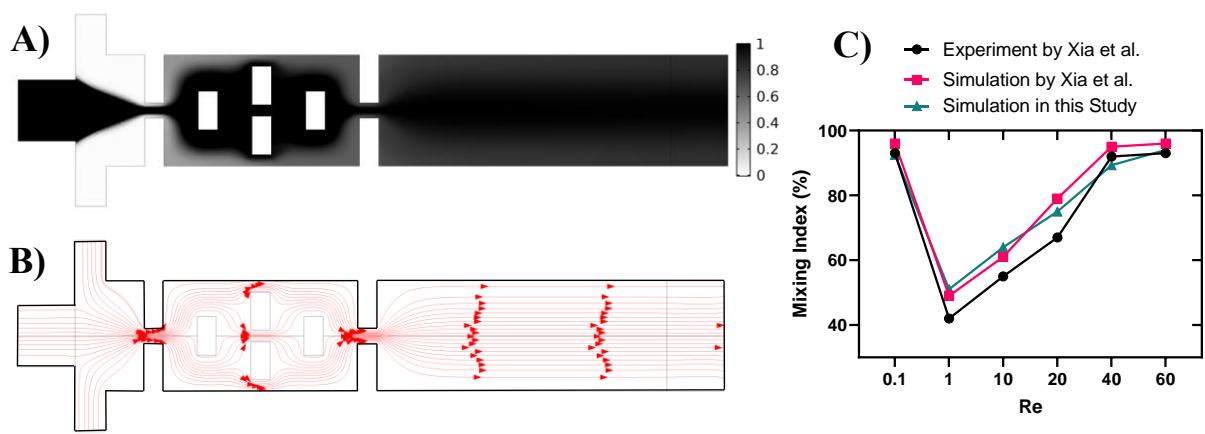

**Figure 2. Model Validation. A)** Dye concentration and **B)** streamlines at $Re = 0.1$ in the geometry of experimental study by Xia et al.[46]. **C)** Comparison of numerical results of mixing indices with results by Xia et al.[46].

# 3. Results and Discussion

## 3.1. Effect of $S_a$

In this section, we investigate the impact of the $S_a$ on both the mixing index and pressure drop, considering a zero gap size (G) and a pillar diameter of 100 $\mu m$. **Fig. 3** illustrates the flow pattern and dye concentration within the channel, showcasing the mixing efficiency for various $S_a$ values in both slanted and arrowhead configurations. The $S_a$ value dictates the flow direction within the pillar array. Within **Figs. 3E-H**, dashed boxes highlight the interfaces between fluids with distinct concentrations. These interfaces play a crucial role in species diffusion and mixing in the channel. Red arrows indicate the flow direction through the pillar array. When $S_a$ has small values, the proximity of the pillar columns together promotes horizontal flow, with lower resistance compared to the direction alongside the pillar columns (**Figs. 3E, G**). Consequently, the interface between fluids with different concentrations gets separated by pillars throughout the channel and recombines, enhancing the mixing index. Conversely, for larger $S_a$ values, the distance between pillar columns is sufficient to balance the resistance in both horizontal and column-aligned directions (**Figs. 3F, H**). This results in fluid flow occurring predominantly alongside the column of pillars, reducing the separation of the interface between the two fluids and leading to a lower mixing index. To be noted that in case of perfect mixing, the color in the mixing box at the end of the channel becomes green with relative concentration of 0.5 (**Figs. 3A-D**).

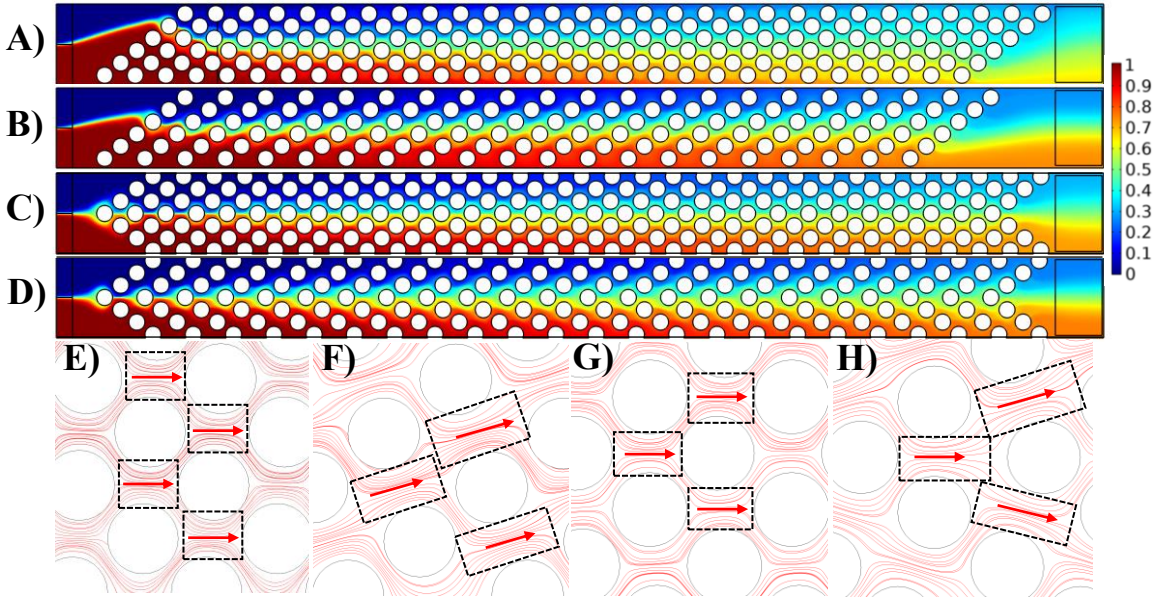

**Figure 3. Demonstration of flow pattern and mixing efficiency with respect to $S_a$**. Dye concentration throughout the channel for the slanted configuration with **A)** $S_a = 190\mu m$, **B)** $S_a = 250\mu m$, and arrowhead configuration with **C)** $S_a = 190\mu m$, and **D)** $S_a = 250\mu m$. Magnified view of the streamlines passing through the pillars in slanted configuration with **E)** $S_a = 190\mu m$, **F)** $S_a = 250\mu m$, and arrowhead configuration with **G)** $S_a = 190\mu m$, and **H)** $S_a = 250\mu m$.

**Fig. 4A** illustrates the quantified comparison of mixing indices between the two configurations concerning $S_a$. The mixing index in the arrowhead configuration is marginally lower than that in the slanted configuration. This discrepancy can be attributed to the flow direction of the two fluids with different concentrations. In the arrowhead configuration (see **Fig. 3D**), the red fluid descends alongside the column of pillars, while the blue fluid ascends. This results in less effective contact between the fluids compared to the slanted configuration. In the slanted configuration (see **Fig. 3B**), the red fluid ascends alongside the row of pillars, effectively mixing with the blue fluid. This elongates the interface between the two fluids, consequently enhancing the mixing index. To be noted in this all sections of this study, the dye concentrations are illustrated and mixing indices are computed once the mixing process reaches steady state at the end of the channel.

Pressure drop is a crucial parameter in the design of micromixers. While incorporating obstacles can enhance mixing efficiency, it also has the potential to elevate the pressure drop. A higher pressure drop necessitates increased pressure and driving force to drive fluids through the channel. This heightened pressure may lead to the microfluidic chip bursting or compromise the bonding of the chip. In **Fig. 4B**, the pressure drop (difference in pressure between the inlet and outlet) for both configurations is depicted in relation to $S_a$. As $S_a$ increases, the distance between pillar columns widens. Consequently, fluid can traverse through the pillars more effortlessly, resulting in a decrease in the pressure drop.

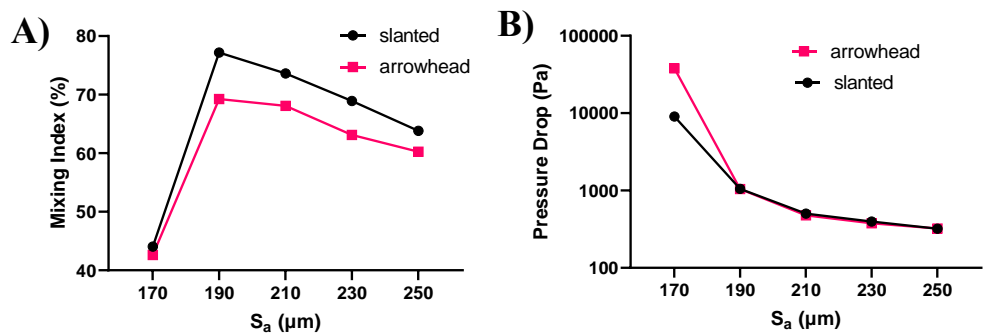

**Figure 4. Mixing index and pressure drop variations with $S_a$. A)** mixing index and **B)** pressure drop.

### 3.2. Effect of Gap Size (G)

In this section, we explore the impact of the gap size (G) between groups of pillars, with a fixed pillar diameter of 100 $\mu m$. **Fig. 5** illustrates the mixing efficiency and flow characteristics within the channel, emphasizing the role of the gap as an analyte distributor (see **Fig. 5D**). In the slanted configuration, the red fluid ascends before each pillar group (green arrow in **Fig. 5D**), distributing itself among various pillar rows (blue arrows). This distribution enhances the diffusion between fluids with distinct concentrations, elongating the interface and thereby increasing the mixing index. Comparing **Fig. 5A and B** highlights the effect of gap size (G) while keeping $S_a$ constant. The comparison reveals that an increase in G results in decreased mixing efficiency. This is attributed to the diminishing resistance in the gap region (green arrow in **Fig. 5D**) as G increases, causing uneven distribution among the rows of pillars (blue arrows). In simpler terms, a substantial portion of the fluid ascends and passes through the last rows of pillars (upper blue arrows). This uneven distribution is evident in **Figs. 5A and B**, where the dye distribution is more uniform in **Fig. 5A** across the pillar rows compared to **Fig. 5B**. Another contributing factor is the degree of contraction and expansion in the gap region. A comparison of **Figs. 5G and H** reveals that smaller gap sizes result in more pronounced contractions and expansions in the gap region, contributing to better mixing. **Fig. 7A** provides a quantified comparison of the mixing index across various $S_a$ and G values for the slanted configuration.

The comparison between **Figs. 5A and C** highlights the influence of gap size (G) under different $S_a$ values. The dye distribution illustrates that, with a constant gap size, increasing $S_a$ results in a decrease in the mixing efficiency. **Figs. 5D and F** demonstrate that as $S_a$ increases, the distance between pillars increases. As a result, fluid can easily flow horizontally through the pillars due to low resistance, and there is not much upward flow in the gap region in **Fig. 5F** compared to **Fig. 5D** (green arrow). Consequently, there is reduced distribution of the fluid among various rows of pillars (blue arrows), leading to a smaller interface for diffusion. In **Fig. 5I**, a magnified view of velocity vectors for $S_a = 250\ \mu m$ and $G = 50\ \mu m$ is presented, clearly depicting that the fluid can effortlessly traverse through the pillars without significant alteration in its direction alongside the pillar columns.

**Fig. 7A** illustrates that the slanted configuration with no gap ($G = 0$) provides the lowest mixing index across all $S_a$ values. In the slanted configuration, as the pillar columns slope downward, the fluids naturally flow in a downward direction. Introducing a gap between groups of pillars reverses this flow, causing the fluid to move upward in the gap and then distributing it downward into the next pillar group. This upward and downward movement significantly increases the mixing interface. In the absence of a gap, as seen in the gap-less system, there is no upward flow in the gap and subsequent distribution, resulting in a lower mixing index (compare **Figs. 3A and 5A**).

**Fig. 7C** illustrates the pressure drop for the slanted configuration at various $S_a$ values with respect to G. As anticipated, an increase in the gap size leads to a reduction in the pressure drop. This is attributed to the lower resistance against the flow, given the increased free space in the channel and a decreased number of pillars.

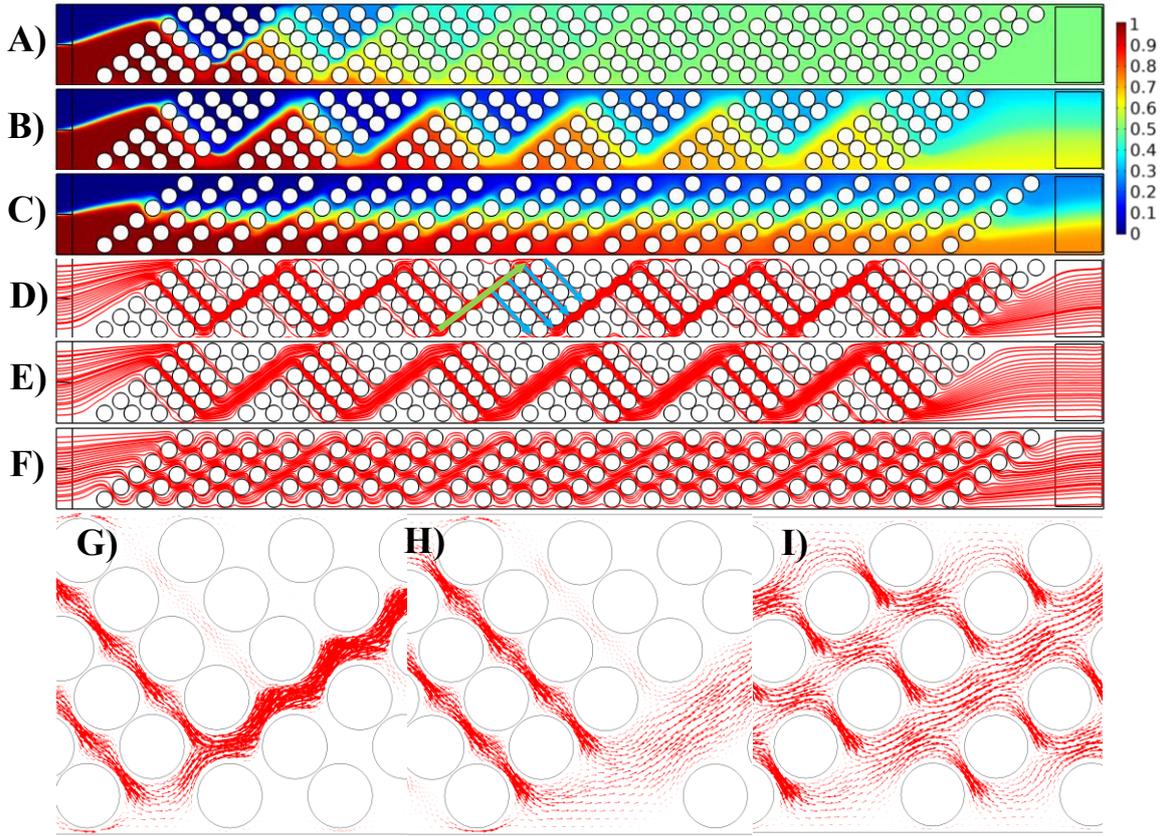

**Figure 5. Demonstration of flow pattern and mixing efficiency with respect to gap size (G) in the slanted configuration**. Dye concentration throughout the channel with **A)** $S_a = 170\ \mu m$ and $G = 50\ \mu m$, **B)** $S_a = 170\ \mu m$ and $G = 200\ \mu m$, and **C)** $S_a = 250\ \mu m$ and $G = 50\ \mu m$. Streamlines passing through the pillars with **D)** $S_a = 170\ \mu m$ and $G = 50\ \mu m$, **E)** $S_a = 170\ \mu m$ and $G = 200\ \mu m$, and **F)** $S_a = 250\ \mu m$ and $G = 50\ \mu m$. Magnified view of the velocity vectors through the pillars for **G)** $S_a = 170\ \mu m$ and $G = 50\ \mu m$, **H)** $S_a = 170\ \mu m$ and $G = 200\ \mu m$, and **I)** $S_a = 250\ \mu m$ and $G = 50\ \mu m$.

In the arrowhead configuration, we observe a similar pattern in the variations of mixing index and pressure drop. **Fig. 6** illustrates the mixing efficiency and flow characteristics within the channel, emphasizing the role of the gap G as an analyte distributor (see **Fig. 6D**) in the arrowhead configuration. In the arrowhead configuration, the interface between the red and blue fluids becomes separated, with part ascending and part descending before each pillar group (indicated by green arrows in **Fig. 6D**). Subsequently, these separated flows are distributed through the pillar rows (depicted by blue arrows in **Fig. 6D**). After this distribution, the separated flows come into contact again after the pillar group and are once more separated in the next gap. This distribution enhances diffusion between fluids with distinct concentrations, elongating the interface and consequently increasing the mixing index. Comparing **Fig. 6A and B** underscores the impact of gap size (G) while maintaining a constant $S_a$. The comparison reveals that an increase in G leads to decreased mixing efficiency. This is attributed to the diminishing resistance in the gap region as G increases, causing uneven distribution among the row of pillars. In simpler terms, a substantial portion of the fluid moves along the pillar rows in the gap and passes through the last rows of pillars (blue arrows near the upper and lower walls of the channel). This uneven distribution is evident in **Figs. 6A and B**, where the dye distribution is more uniform in **Fig. 6A** across the pillar rows compared to **Fig. 6B**. Another contributing factor is the degree of contraction and expansion in the gap region. A comparison of **Figs. 6G and H** reveals that smaller gap sizes result in more pronounced contractions and expansions in the gap region, contributing to better mixing. **Fig. 7B** provides a quantified comparison of the mixing index across various $S_a$ and G values for the arrowhead configuration.

The comparison between **Figs. 6A and C** sheds light on the impact of gap size (G) under different $S_a$ values in the arrowhead configuration. The dye distribution reveals that, with a constant gap size, an increase in $S_a$ results in a decrease in mixing efficiency. **Figs. 6D and F** demonstrate that as $S_a$ increases, the distance between pillars widens. Consequently, the fluid can easily flow horizontally through the pillars due to low resistance, resulting in minimal upward and downward flow in the gap region in **Fig. 6F** compared to **Fig. 6D** (indicated by green arrows). Consequently, there is diminished distribution of the fluid among various rows of pillars (depicted by blue arrows), leading to a smaller interface for diffusion. In **Fig. 6I**, a magnified view of velocity vectors for $S_a = 250\ \mu m$ and $G = 50\ \mu m$ is presented, clearly depicting that the fluid can effortlessly traverse through the pillars without significant alteration in its direction alongside the pillar rows.

**Fig. 7B** illustrates that the arrowhead configuration with no gap ($G = 0$) yields the lowest mixing index across all $S_a$ values. In the arrowhead configuration, the pillar columns slope downward in the upper half of the channel and slope upward in the lower half. Consequently, the fluids naturally converge toward the middle of the channel, resulting in a short interface between the two fluids solely in the middle of the channel, leading to low mixing efficiency (**Fig. 3C**). Introducing a gap between groups of pillars reverses this flow, causing the fluid to move upward and downward toward the walls of the channel (green arrows in **Fig. 6D**) and then distributing it into the next pillar group toward the middle of the channel (blue arrows in **Fig. 6D**). This upward and downward movement significantly increases the mixing interface. In the absence of a gap, as observed in the gap-less system, there is no upward and downward flow in the gap and subsequent distribution, resulting in a lower mixing index (compare **Figs. 3C and 6A**).

**Fig. 7D** illustrates the pressure drop for the arrowhead configuration at various $S_a$ values with respect to G. As anticipated, an increase in gap size leads to a reduction in the pressure drop. This is attributed to the lower resistance against the flow, given the increased free space in the channel and a decreased number of pillars.

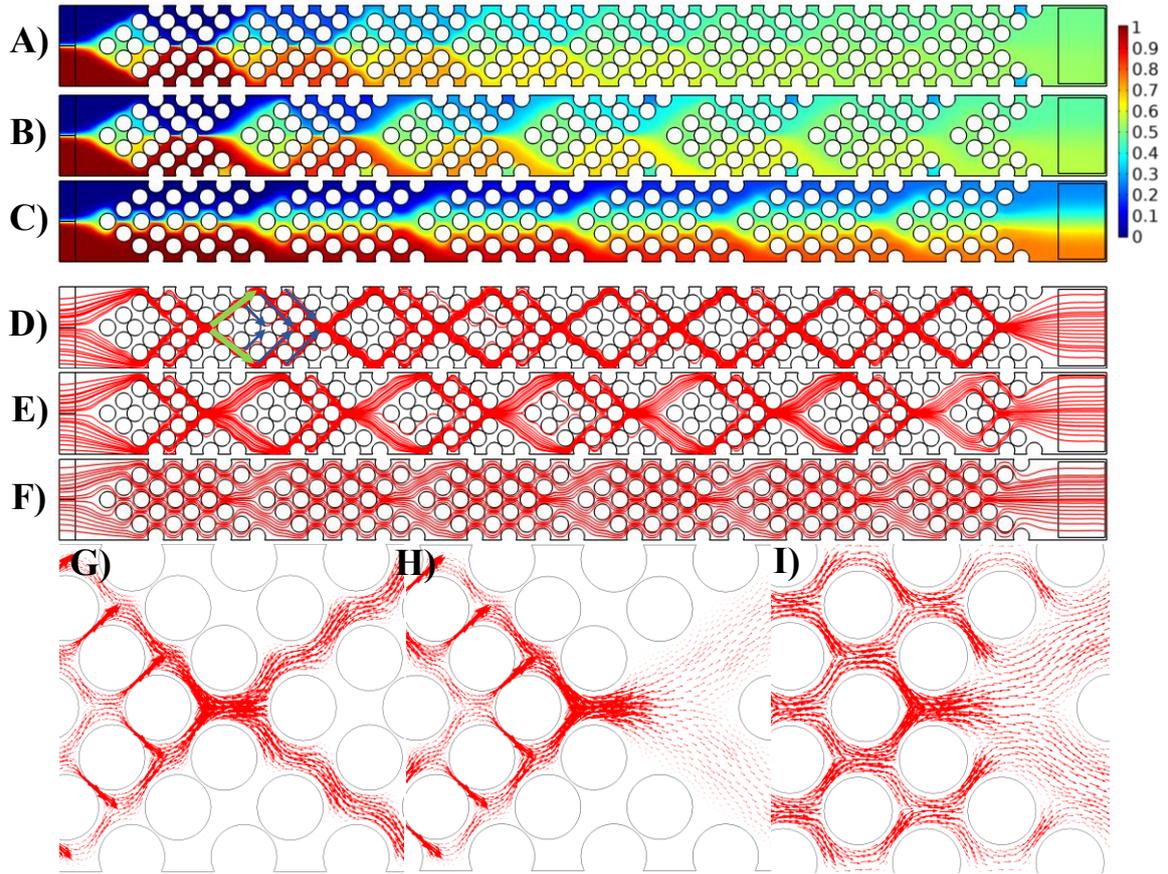

**Figure 6. Demonstration of flow pattern and mixing efficiency with respect to gap size (G) in the arrowhead configuration**. Dye concentration throughout the channel with **A)** $S_a = 170\ \mu m$ and $G = 50\ \mu m$, **B)** $S_a = 170\ \mu m$ and $G = 200\ \mu m$, and **C)** $S_a = 250\ \mu m$ and $G = 50\ \mu m$. Streamlines passing through the pillars with **D)** $S_a = 170\ \mu m$ and $G = 50\ \mu m$, **E)** $S_a = 170\ \mu m$ and $G = 200\ \mu m$, and **F)** $S_a = 250\ \mu m$ and $G = 50\ \mu m$. Magnified view of the velocity vectors through the pillars for **G)** $S_a = 170\ \mu m$ and $G = 50\ \mu m$, **H)** $S_a = 170\ \mu m$ and $G = 200\ \mu m$, and **I)** $S_a = 250\ \mu m$ and $G = 50\ \mu m$.

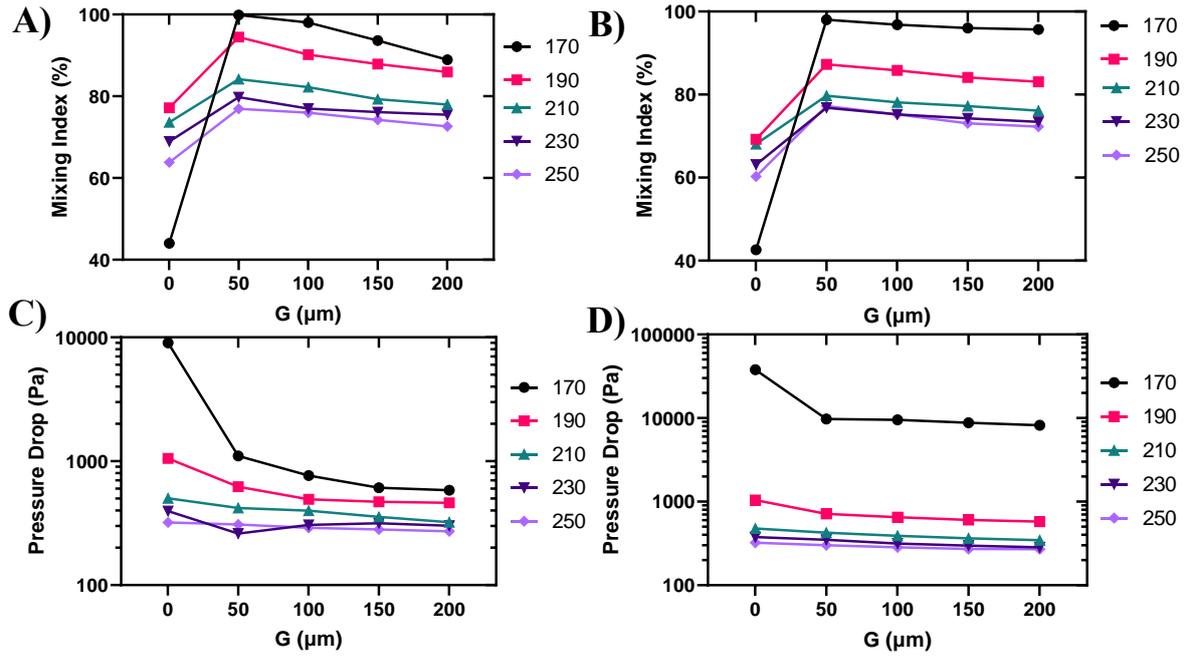

**Figure 7. Mixing index and pressure drop variations with gap size (G).** Mixing index in **A)** slanted configuration and **B)** arrowhead configuration. Pressure drop in **C)** slanted configuration and **D)** arrowhead configuration.

### 3.3. Effect of Pillar Diameter

In this section, we examine the impact of pillar diameter on both the mixing index and pressure drop. Building upon insights gained from previous sections, where the pillar array with $S_a = 170\ \mu m$ and $G = 50\ \mu m$ exhibited the highest mixing index, we select these parameters for further investigation into pillar diameter effects. To ensure a fair comparison across different pillar diameters and isolate the impact of diameter variation, we maintain constant distances between pillars within each column and between adjacent columns of pillars. This investigation is conducted within the slanted configuration, with $S_a = 170\ \mu m$ and $G = 50\ \mu m$. **Fig. 8A** displays the variations in mixing index with respect to pillar diameter. As the pillar diameter increases from $50\ \mu m$ to $125\ \mu m$, the mixing index increases, as well (compare **Figs. 9C, D**). This phenomenon can be attributed to the larger pillars providing more space within the passageways between pillar rows, for a constant distance between pillars in each column and between the adjacent columns. A comparison between **Figs. 9E and F** reveals that with $D_p = 125\ \mu m$, there is greater expansion between pillars (indicated by the black dashed box) compared to $D_p = 50\ \mu m$. Since the distances between pillars remain constant across cases with different diameters (indicated by the green dashed box), the size disparity between contraction (green dashed box) and expansion (black dashed box) increases with larger pillar diameter, leading to improved mixing and higher mixing index values. Furthermore, in cases with smaller pillar diameters, the overall space between rows of pillars (sum of contraction and expansions) is reduced, resulting in higher flow velocity and shorter residence time for analytes to mix (compare **Figs. 9A, B**). Consequently, this contributes to a lower mixing index. To be noted that the number of pillar columns in each pillar group changes with pillar diameter so that we have a constant pillar group length.

**Fig. 8B** presents the variations in pressure drop with pillar diameter. The trend indicates that as the pillar diameter increases, the pressure drop initially decreases before increasing again. Notably, the pillar array with a diameter of $100\ \mu m$ exhibits the lowest pressure drop while achieving the second highest mixing index (**Fig. 8A**). Comparing our findings with similar studies conducted by other researchers, our results demonstrate a reasonable pressure drop alongside a considerable improvement in the mixing index. For instance, in the work by Xia et al.[46], they employed micromixers featuring contraction-expansion and obstacles to enhance mixing. In their study, at $Re = 1$, the pressure drop was approximately $1000\ Pa$, with a mixing index of about 48%. In contrast, in our study, we were able to maintain a pressure drop of around $1000\ Pa$ while significantly increasing the mixing index to more than 98% (**Fig. 8A**). This suggests that our approach yields superior mixing efficiency without substantially increasing the pressure drop, which is a noteworthy advancement in microfluidic micromixer design.

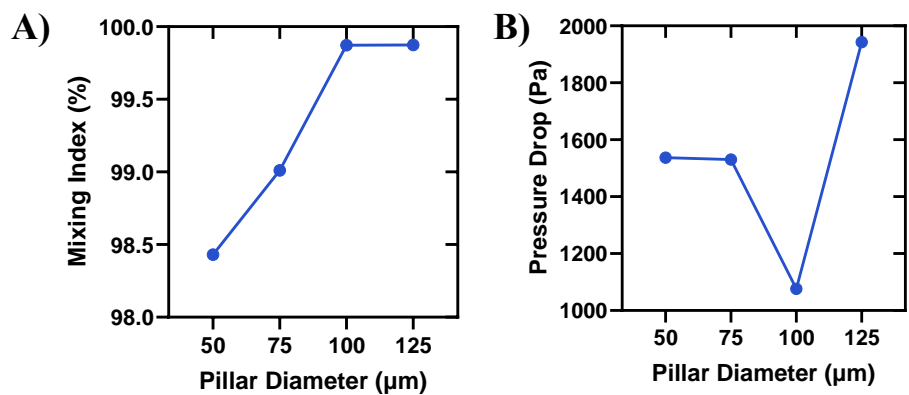

**Figure 8. Mixing index and pressure drop variations with pillar diameter ($D_p$). A)** Mixing index and **B)** pressure drop.

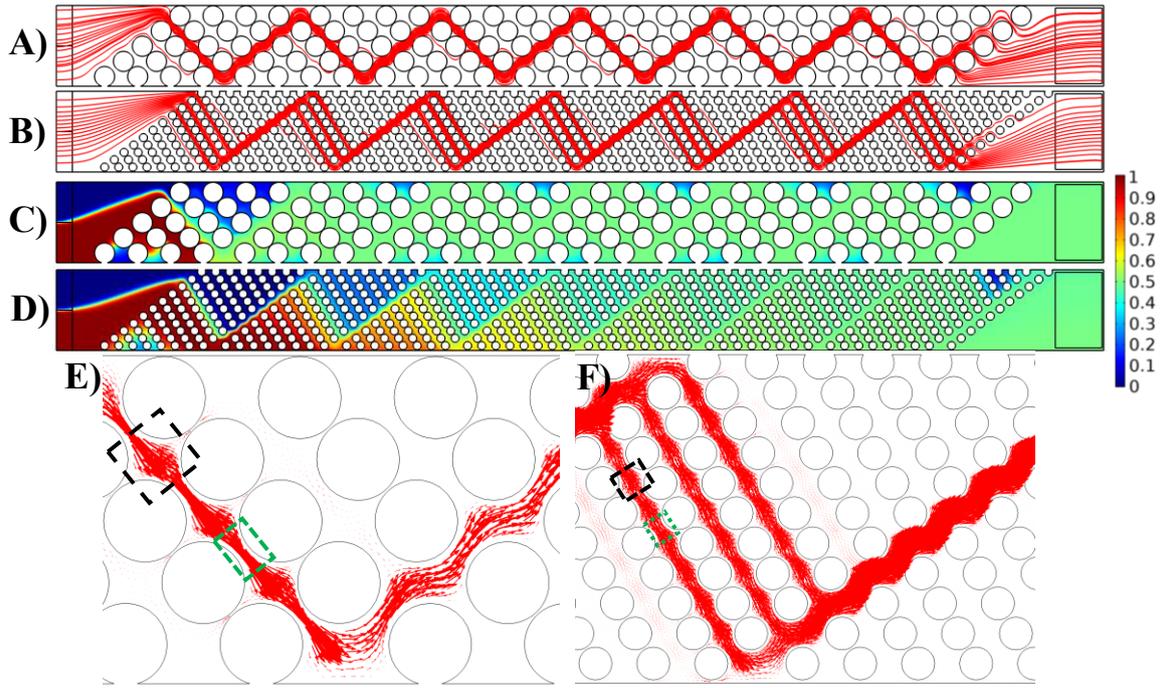

**Figure 9. Demonstration of flow pattern and mixing efficiency with respect to pillar diameter ($D_p$) in the slanted configuration**. Streamlines passing through the pillars for **A)** $D_p = 125\ \mu m$ and **B)** $D_p = 50\ \mu m$. Dye concentration for **C)** $D_p = 125\ \mu m$ and **D)** $D_p = 50\ \mu m$. Magnified view of the velocity vectors through the pillars for **E)** $D_p = 125\ \mu m$ and **F)** $D_p = 50\ \mu m$.

### 3.4. Effect of vertical shift ($S_y$) of Pillar groups

One of the factors influencing the mixing efficiency in this channel is the vertical shift ($S_y$) between the groups of pillars. Based on the results from previous sections, we chose the best case in terms of mixing index and pressure drop with $S_a = 170\ \mu m$, $G = 50\ \mu m$, and $D_p = 100\ \mu m$. Having a pillar array with these parameters, we changed $S_y$ from 0 to 40 $\mu m$ and investigated the variations in the mixing index and pressure drop. As illustrated in **Fig. 10A**, in the slanted configuration, increasing $S_y$ from 0 to 40 $\mu m$ causes a 4% decrease in the mixing index. Conversely, in the arrowhead configuration, augmenting $S_y$ initially leads to an increase in the mixing index, followed by a slight decline. The maximum mixing index for this configuration is achieved at $S_y = 10\ \mu m$.

The changes in pressure drop with $S_y$ are illustrated in **Fig. 10B**. As the vertical shift increases, the pressure drop decreases. This is due to the fact that with an increasing vertical shift, the rows of pillars in adjacent groups are not directly aligned with each other (see **Fig. 1B**). Consequently, this reduces the resistance to flow, resulting in a decrease in the pressure drop.

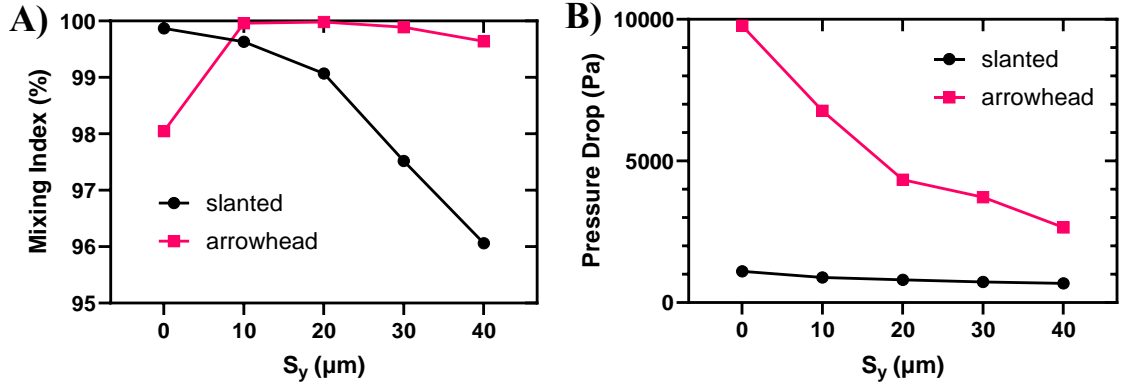

**Figure 10. Mixing index and pressure drop variations with vertical shift ($S_y$). A)** Mixing index and **B)** pressure drop.

### 3.5. Changes in Mixing Index over Time and Channel Length

Based on the parameters investigated and the results obtained in the previous sections, the optimal micromixers in terms of mixing index are as follows: slanted configuration with $S_a = 170\ \mu m$, $G = 50\ \mu m$, $D_p = 100\ \mu m$, and $S_y = 0\ \mu m$; and arrowhead configuration with $S_a = 170\ \mu m$, $G = 50\ \mu m$, $D_p = 100\ \mu m$, and $S_y = 10\ \mu m$ (**Fig. 10A**). Now, for these micromixers, we examine the variations in mixing index over time until reaching steady state at the end of the channel, within the mixing box (**Fig. 1A**). Additionally, exploring the mixing index variations along the channel from the inlet to the outlet provides valuable insight into the mixing length.

**Fig. 11A** illustrates the changes in mixing index over time. The mixing index is calculated within the mixing box at the end of the channel. As time progresses, the mixing index increases until the mixing process reaches a steady state. By the 8-second mark, we achieve a steady state at the end of the channel with a mixing index of 99.5%. In the steady state, both the slanted and arrowhead configurations exhibit high mixing indices near 100%, which is a significant achievement. By the 3-second mark, both configurations can attain mixing indices exceeding 90%.

Another crucial parameter in micromixers is the mixing length, which determines the overall length of the channel over which perfect mixing can be achieved. **Fig. 11B** demonstrates the mixing index variations in different positions along the channel. In both configurations, mixing indices higher than 99% can be achieved with a length of $4300\ \mu m$. In similar studies, such as that by Xia et al.[46], for Re around 1, the reported mixing index is approximately 48% with a channel length of $1600\ \mu m$. In our study, the mixing indices at a distance of $1600\ \mu m$ from the inlet are 54% and 75% for the slanted and arrowhead configurations, respectively.

As a result, both the slanted and arrowhead configurations with the optimum parameters introduced in this section represent ideal micromixers, offering high mixing indices in a short time and with a short mixing length. It is important to note that micromixers typically achieve high mixing indices close to 100% at extremely low ($Re \ll 1$) or high Reynolds Numbers ($Re > 40$)[46]. This phenomenon occurs because at low Re, the residence time increases, allowing analytes more time to mix. Conversely, at high Re, separated vortices can form, enhancing mixing efficiency. However, at moderate Re, the residence time is shorter compared to low Re, and the absence of separated vortices like those at high Re results in poor diffusion and convection, leading to lower mixing efficiency. However, the micromixers introduced in this study demonstrate mixing indices higher than 99.7% at $Re \approx 1$.

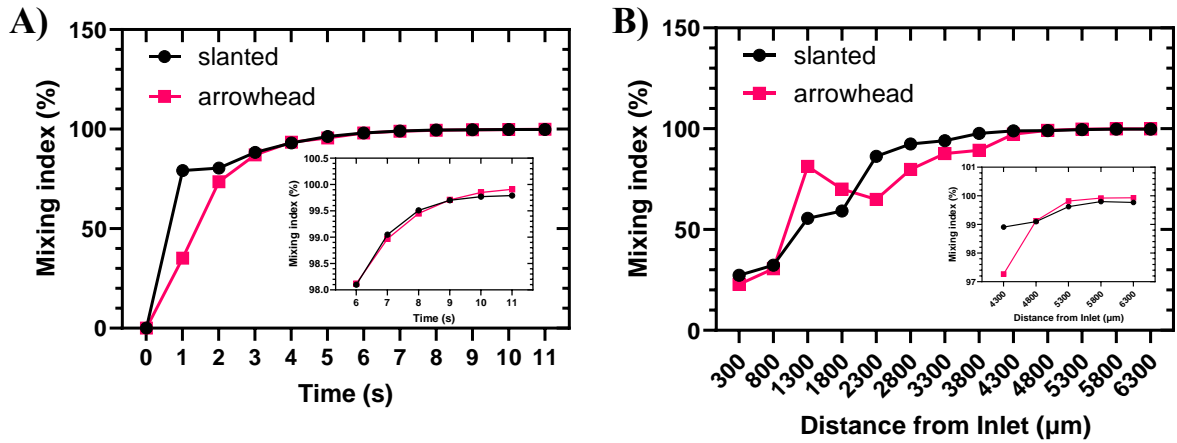

**Figure 11. Mixing index variations with time and distance from inlets. A)** Time-dependent variations of mixing index until reaching steady-state. **B)** Mixing index variations at different positions along the channel.

## 4. Conclusions

This study employed numerical simulations to explore the fluid dynamics and mass transfer characteristics of two innovative micromixers featuring pillar arrays. Investigating how the design of these pillar arrays influences mixing efficiency and pressure drop, we utilized principles such as contraction-expansion and split-recombine. Introducing two distinct configurations of pillar arrays, slanted and arrowhead, we examined parameters including pillar diameter, gap size between pillar groups, distance between pillars, and vertical shift between pillar groups. Through this comprehensive analysis, we identified optimal micromixers capable of achieving exceptional mixing efficiency exceeding 99.7% at moderate Reynolds number ($Re = 1$). This is while at this order of Re achieving high mixing index is challenging. Notably, the pressure drop remained minimal at 1102 Pa. Additionally, our study examined variations in mixing index over time and along different positions within the channel. Both configurations exhibited short mixing lengths and times. We achieved steady state with a 99.5% mixing index at the end of the channel by 8 seconds. Furthermore, by 3 seconds, both configurations surpassed 90% mixing indices. At a distance of 4300 $\mu m$ from the inlet, both configurations achieved mixing indices higher than 99%. At a distance of 1300 $\mu m$, the mixing indices were 53% and 80% for the slanted and arrowhead configurations, respectively. Overall, the swift mixing, minimal pressure drop, and short mixing lengths observed in these novel micromixers position them as promising options for microfluidic applications.